# Exceptional point-based ultrasensitive surface acoustic wave gas sensor

Xingyu Lu[1,#], Yang Yuan[2,#], Fa Chen[2], Xiaoxiao Hou[2], Yanlong Guo[1], Leonhard Reindl[3], Wei Luo[2,✉] and Degang Zhao[1,✉]

[1]Department of Physics, Huazhong University of Science and Technology.
[2]School of Integrated Circuits, Huazhong University of Science and Technology.
[3]Institute for Microsystem Technology, Faculty for Engineering, University of Freiburg, Germany.
[#]These authors contributed equally.
✉email: luowei@hust.edu.cn; dgzhao@hust.edu.cn.

Exceptional points (EPs) refer to degeneracies in non-Hermitian systems where two or more eigenvalues and their corresponding eigenvectors coalesce. Recently, there has been growing interest in harnessing EPs to enhance the responsivity of sensors. Significant improvements in the sensitivity of sensors in optics and electronics have been developed. In this work, we present a novel ultrasensitive surface acoustic wave (SAW) gas sensor based on EP. We demonstrate its ability to significantly respond to trace amount of hydrogen sulfide ( $H_2S$ ) gas by tuning additional loss to approach the EP, thereby enhancing the responsivity compared to the conventional delay line gas sensors. In addition to high sensitivity, our sensor is robust to temperature variation and exclusive to $H_2S$ gas. We propose an innovative method for designing a new generation of ultrasensitive gas sensor.

In non-Hermitian physical systems, exceptional points (EPs) are singularities in the parameter space where eigenvalues and their corresponding eigenvectors coalesce. Systems exhibiting parity-time (PT) symmetry provide an excellent platform for studying EPs[1-3], and significant advances have been achieved in fields such as photonics[4-7], electronics[8] and acoustics[9-11]. EPs have garnered extensive attention for their great potential to enhance the responsivity of sensors[12-20]. This peculiar topological sensitive property is that in the proximity of an $N$th order ($N$ = 2, 3, 4...) EPs, small perturbation $\varepsilon$ will lead to frequency detuning: $\Delta\omega = |\omega - \omega_{EP}| \propto \varepsilon^{1/N}$. Even in the proximity of the second-order EP, the sensor exhibits much greater sensitivity than the linear frequency response under small perturbations. Therefore, leveraging this characteristic of EPs has been theoretically explored to enhance sensing performance[21-23]. Several studies have successfully translated the theoretical predictions to some practical sensors[24], most of which are based on optical microcavities[15,25,26] and electric circuits[12,13,27]. There is a notable scarcity of sensors based on EPs that have been constructed in the acoustic system.

Surface acoustic wave (SAW) sensors have broad application prospects owing to their compact size, digital output, high sensitivity, and ease for wireless passive application[28,29]. Researchers have integrated diverse gas sensitive materials onto the piezoelectric substrate to design efficient SAW gas sensors[30], and one of the major categories, whose physical sensing mechanism mainly relies on the acoustoelectric effect[31]. When the detected gas is physically adsorbed by the gas sensitive materials, it alters the concentration of charge carrier of the piezoelectric substrate and subsequently induces the change of acoustic wave velocity and wave attenuation, ultimately results in a frequency shift in transmission spectrum. Mostly, researchers focus on the center frequency shift and often regard the associated loss from the acoustoelectric effect as the obstacle to achieve highly sensitive sensors. However, the introduced dissipation from gas sensitive materials naturally forms a divergent non-Hermitian system, which has the potential to generate EP to enhance sensitivity.

In this study, based on a passive PT-symmetric system comprising two coupled resonators defined by Bragg mirrors, we present the experimental demonstration of a SAW gas sensor with significantly enhanced sensitivity to detect trace amount of hydrogen sulfide ($H_2S$) gas effectively. A $SnO_2$ thin film, which is sensitive to $H_2S$ gas, is deposited onto one of the resonators. As $H_2S$ is adsorbed by $SnO_2$ thin film, additional loss is introduced to the system. By carefully managing the additional loss to render the system in the proximity of EP, a small perturbation in the form of loss is deliberately introduced, resulting in extremely obvious frequency splitting in the transmission peak. Notably, that frequency splitting scales proportionally to the square root of the perturbation strength, which is more sensitive than conventional linear response gas sensor. Furthermore, our sensor is of excellent performance in sensing, such as rapid response to little gas, robustness to temperature variations, exclusivity in $H_2S$ gas detection and good recoverability. Our ultrasensitive sensor represents a departure from conventional SAW gas sensors, offers a novel sensing approach and proposes pioneering scheme applicable to gas sensing systems (but not limited to), thereby opening up a new perspective for the development of the next-generation sensors.

**Experimental setup**

To date, some efforts have been made to introduce gain to balance spontaneous loss to construct PT-symmetric systems, including tuning the power of the pump laser in photonics[6,7,32], constructing equivalent negative resistance in electronics[8,12,13] and controlling speakers with an active circuit in acoustics[33,34]. In SAW system, loss is natural to material, but introducing and balancing gain and loss are not easy to implement. To address this, we adopt a model of two coupled resonators, forming a passive PT-symmetric system[35], which only requires precise control the loss instead of introducing gain, as shown in Fig. 1a. Given the robust electromechanical coupling and low propagation loss, our device is constituted on a 128° Y-cut lithium niobite ($LiNbO_3$) substrate. The white gratings represent the input and output interdigital transducers (IDTs) with 60 aluminum finger electrodes each. IDTs serve as the wave source and detector, which enable the mutual conversion of electrical signals and surface acoustic wave signals. Two coupled resonators are defined by three spaced reflecting gratings (Bragg mirrors). The coupling strength $\kappa$ between the two resonators can be tuned by adjusting the number of electrodes in the middle Bragg mirror[35]. For better coupling, the electrodes in the gratings and IDTs have identical width and spacing. The pink plane represents a $H_2S$ gas sensitive $SnO_2$ thin film deposited onto one resonator. After injecting a certain amount of $H_2S$ gas, with time evolution, the $H_2S$ gas is continuously adsorbed by the $SnO_2$ thin film and provide extra electronics. The increase of charge carrier concentration in semiconductor thin films results in a gradual increase of additional loss $\gamma$ due to the interaction between lower electron transport speeds and surface acoustic waves, allowing the state of system to approach the EP ($\gamma = 2\kappa$). The analytical eigensolutions of our two coupled resonators system can be found in Supplementary Materials S1. It is important to note that the deposited $SnO_2$ thin film will slightly alter the resonant frequency of resonator, leading to $\omega_1 \neq \omega_2$. It will disastrously cause the annihilation of EP and induce a baseline bifurcation of eigenfrequency[1,12]. To overcome that, we introduce additional number of electrodes in the rightmost Bragg mirror to compensate the frequency deviation caused by thin film. After optimization, the appropriate numbers of electrodes of left (near input IDT), middle, and right (near output IDT) Bragg mirrors are 40, 60, and 50, respectively. The details about optimization can be found in Supplementary Materials S4.

The photograph of the actual experimental setup is shown in Fig.1b, including a 1L reaction chamber, a printed circuit board (PCB) and a SAW chip carrying $SnO_2$ thin film. The SAW device and the measuring circuit are assembled onto the PCB, which is placed in the reaction chamber equipped with two gas valves for gas injection and evacuation. During the experiment, a network analyzer is connected to the electrodes on the chamber wall to measure the transmission spectrum. When there is no gas injection, the additional loss is zero and the system is in the "PT exact phase" regime[6]. Hence the different real parts of eigenfrequencies result in transmission-peak splitting in the experimentally observed transmission spectrum (upper panel of Fig. 1c). However, when we inject 0.4ppm of $H_2S$ gas into the reaction chamber, as time goes by, the two transmission peaks merge into a single peak until the system stabilizes in the "PT-broken phase" regime, as is shown in the lower panel of Fig. 1c. Now the two eigenfrequencies share identical real parts, leading to only one resonance peak in the transmission spectrum.

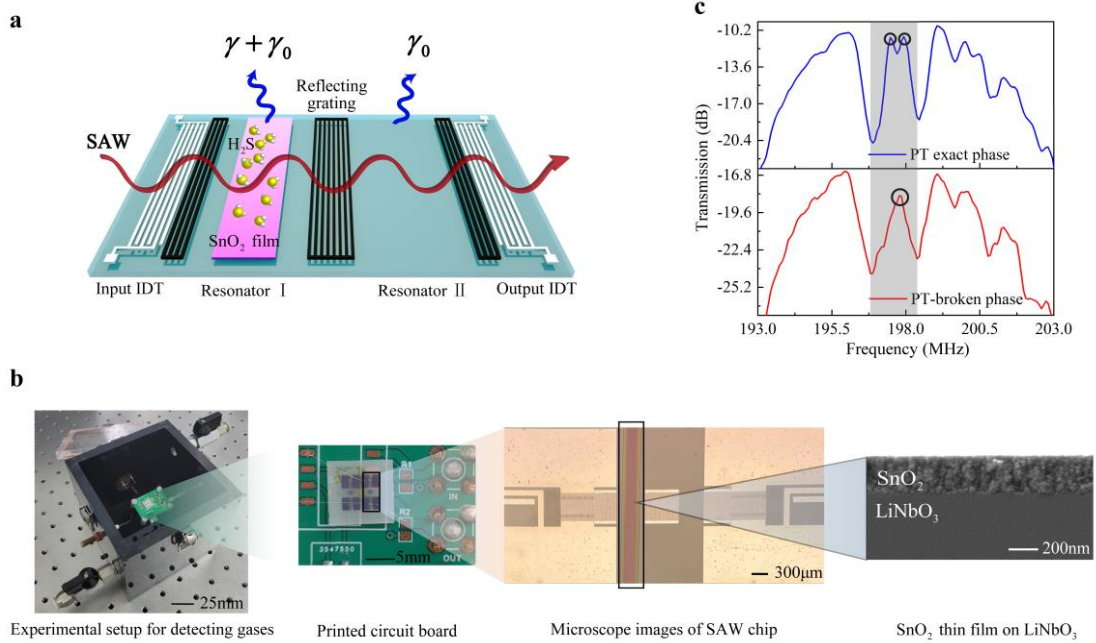

**Fig. 1 | The passive PT-symmetric SAW gas sensor. a**, Schematic diagram of the passive PT-symmetric SAW gas sensor. The two coupled resonators are defined by three Bragg mirrors and coupled to the SAW transmission line. A $SnO_2$ thin film is deposited onto the resonator I. $\gamma_0$ denotes the inherent loss of the resonators. The injection of $H_2S$ can be absorbed by $SnO_2$ thin film and bring about additional loss $\gamma$ in the resonator I. **b**, The actual experimental setups. **c**, Measured transmission spectrum of the SAW gas sensor in the PT exact phase regime (without $H_2S$) and the PT-broken phase regime (with $H_2S$ and additional loss $\gamma$ is significant). Shaded gray region denotes the bandgap of Bragg mirrors.

**Determination of device parameters**

The key to our sensor's effective working lies in tuning the additional loss to ensure the resonant detuning occurs in the proximity of the transmission peak degeneracies[27]. To achieve it, the configuration of our device should be carefully designed. Here, we provide a detailed description about the determination of structural parameters through pre-experiment and numerical fitting of experimental results. The deposited thin film brings about additional loss due to the unequal surface acoustic wave velocity in $LiNbO_3$ substrate and electron drift velocity in $SnO_2$ thin film[37]. A straightforward equivalent RC circuit model can be used to describe the attenuation of a SAW with respect to the wave intensity $I = I_0 e^{-\Gamma k x}$ along the propagation direction[38]:

$$\Gamma = \frac{K^2}{2} \frac{\frac{\sigma_{0r}}{\sigma_0}}{1+(\frac{\sigma_{0r}}{\sigma_0})^2}, \tag{1}$$

where $K^2 = 5.5\%$ is the electromechanical coupling coefficient for 128° Y-cut $LiNbO_3$ substrate, $\sigma_0 = 1.67 \times 10^{-6}/\Omega$ is the sheet conductivity of $LiNbO_3$, $\sigma_{0r}$ is the sheet conductivity of $SnO_2$ thin film, which exhibits sensitive response to $H_2S$ gas. Therefore, by injecting $H_2S$ gas into the system, we can tune the attenuation of the SAW in system through tuning the sheet conductivity of $SnO_2$ according to Eq. (1). It should be noted that the SAW attenuation $\Gamma$ is not the additional loss $\gamma$ in Hamiltonian (Eq. (S1) in Supplementary Materials). The relation between them is $\gamma = 31.5346\Gamma + 0.09166\,(MHz)$, which can be determined by pre-experiment of the individual thin film and numerical fitting of pre-experimental results (details can be found in Supplementary Materials S3).

As depicted in Fig. 2a, $\gamma$ increases with the increase of $\sigma_{0r}/\sigma_0$ and reaches its maximum value $0.525\,MHz$ when $\sigma_{0r}/\sigma_0 = 1$, and then gradually decreases. In the experiment, the necessary condition to observe two-peaks-to-one-peak process is that $\gamma$ must gradually increase from less than $\gamma'$ to greater than $\gamma'$, where $\gamma'$ is the critical value corresponding to the coalescence point of transmission peaks. The deposited thin film through magnetron sputtering technique will bring initial sheet conductivity to the system. That initial state of the thin film predominantly depends on the sputtering time (the other experiment parameters, such as pressure and temperature remain unchanged), and we carry on pre-experiment to determine the appropriate sputtering time. In Fig. 2b, we measure the sheet conductivity of the $SnO_2$ thin film with 30, 45 and 60 minutes of sputtering time. It clearly embodies that the longer the sputtering time, the lower the initial sheet conductivity of $\sigma_{0r}$. After injecting 1ppm $H_2S$ gas, sheet conductivities of all the thin films increase rapidly, which displays the sensitiveness of $SnO_2$ thin film to $H_2S$ gas. If the initial sheet conductivity is too large, the variation region for additional loss will be compressed. Conversely, if the initial sheet conductivity is too small, the reaction time for detecting tiny concentration of $H_2S$ gas will be long. To balance them, we select the 45 minutes sputtering time sample, where the initial $\sigma_{0r}/\sigma_0$ is 0.124 and initial additional loss is 0.199MHz, to carry on the sensing experiment.

Next, it is crucial to determine the critical value $\gamma'$. Firstly $\gamma'$ cannot be greater than the maximum $\gamma$ (0.525MHz). In addition, $\gamma'$ cannot be too small because it must be greater than the initial loss brought by the $SnO_2$ thin film without injection of $H_2S$ gas. Without changing the Bragg mirrors on both sides, $\gamma'$ can be tuned by the number of electrodes in the middle Bragg mirror of our two coupled resonators system[30]. We build up a simulation model to calculate the transmission spectrum with the increase of additional loss. Through carefully tuning the number of electrodes of the middle Bragg mirror, the explicit two-peaks-to-one-peak process has been observed in transmission spectrum, as is shown in Fig. 2c. Now the optimal number of electrodes of the middle Bragg mirror is 60, and the additional loss at critical state is identified to be $\gamma' = 0.393MHz$. We fabricate the sample according to the simulation determined structural parameters and measure the transmission spectrum with the injection of 0.4ppm $H_2S$ gas. The evolution of transmission peaks (Fig. 1c) agrees well with the numerical results (Fig. 2c).

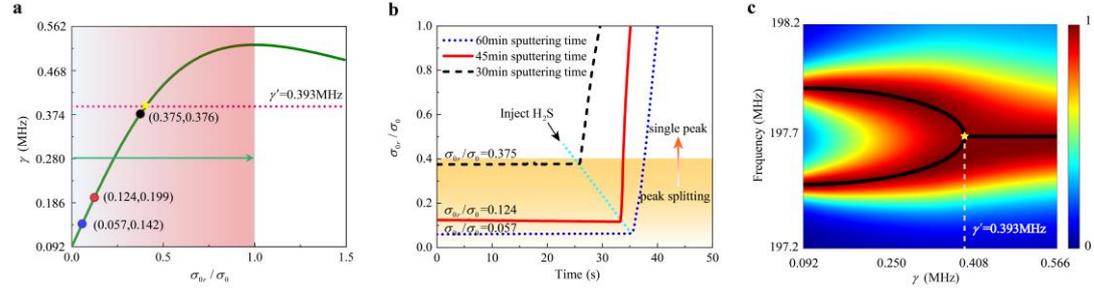

**Fig. 2 | The analysis of additional loss in thin film**, The evolution process of the additional loss with the increase of sheet conductivity ratio $\sigma_{0r}/\sigma_0$. The marked points represent the initial states of three thin films presented in **b**. **b**, For different magnetron sputtering time, $SnO_2$ thin films have different stable initial $\sigma_{0r}/\sigma_0$. After injecting $H_2S$ gas, $\sigma_{0r}/\sigma_0$ of all the thin films increases rapidly. **c**, The transmission spectrum of our SAW gas sensor with the increase of $\gamma$. The black solid curve denotes the position of transmission peaks. The yellow star is the coalescence point of transmission peaks.

**Sensitivity Analysis**

In order to facilitate the observation of two-peaks-to-one-peak transition process in experiment, we record the evolution of transmission spectrum over time with the injection of 0.4ppm $H_2S$. Applying temporal coupled mode theory, the transmissivity can be expressed as[27,39,40] (Supplementary Materials S2):

$$T(\omega) = |S_{21}|^2 = \left| \frac{\kappa \gamma_c}{\kappa^2 + (\gamma_0 + \gamma_c + \gamma - i\Delta)(\gamma_0 + \gamma_c - i\Delta)} \right|^2 \quad (2)$$

where, $\Delta = \omega - \omega_0$ denotes the frequency detuning, $\gamma_c$ is the coupling strength between the coupled resonators system and the transmission line, $\gamma_0$ is the symmetric inherent loss in two resonators. As is shown in Fig. 3a, with weak additional loss $\gamma < \gamma'$, we can observe two separated peaks in transmission spectrum. As time goes by, the gradually increasing $\gamma$ leads to an overlap of the resonant supermodes, consequently, the split peaks gradually merge to a single peak, which exhibit excellent agreement between experiment and theory.

The most common sensing strategy of conventional SAW gas sensors is measuring the center frequency shift caused by disturbance from target gas. To facilitate a comparison with conventional gas sensors and highlight the advantages of our sensor, we also design a delay line gas sensor, which only removes all the Bragg mirrors from our passive PT-symmetric SAW gas sensor (the schematic diagram can be seen in Fig. S7a in Supplementary Materials). Both sensors have deposited $SnO_2$ thin films of the same width and thickness at the same location in order to introduce same disturbance in the environment with the same concentration of $H_2S$ gas. As shown in Fig. 3b, we record the center frequencies of one transmission peak for the delay line gas sensor (Supplementary materials S5) as a reference. Apparently, in the proximity of the coalescence point of transmission peaks, the frequencies of transmission peaks of our passive PT-symmetric sensor dramatically change in a very short time. In contrast, the frequency drift over time of the delay line gas sensor is quite small, making the measurement inaccurate and challenging.

By solving $dT/d\omega = 0$, the frequencies of transmission peaks can be determined as:

$$\omega_{\pm} = \begin{cases} \omega_0 \pm \sqrt{\kappa^2 - \frac{\gamma^2}{2} - \gamma(\gamma_0 + \gamma_c) - (\gamma_0 + \gamma_c)^2}, & for \ \gamma \leq \gamma' \\ \omega_0, & for \ \gamma > \gamma' \end{cases} \quad (3)$$

From Eq. (3), the critical state between two-peaks state and single-peak state is at $\gamma = \gamma'$, where $\gamma'$ is the solution of $\kappa^2 - \frac{\gamma^2}{2} - \gamma(\gamma_0 + \gamma_c) - (\gamma_0 + \gamma_c)^2 = 0$. It must be pointed out that this coalescence point of transmission peaks is not EP, i.e. $\gamma' \neq 2\kappa$ [27,36]. This discrepancy between transmission-peak splitting and eigenvalue splitting is due to the coupling between the coupled resonators system and the transmission line. Then, the appearance of a single transmission peak does not imply the system is certainly in the "PT-broken phase" regime. The detailed analysis about transmissivity can be found in Supplementary Materials S2. In $\gamma \leq \gamma'$ regime, when $\gamma$ is in the proximity of $\gamma'$, we define $\varepsilon=|\gamma-\gamma'|$ as a perturbation and the frequency splitting of the two peaks is $\Delta\omega \approx 2\sqrt{(\gamma'+\gamma_0+\gamma_c)\varepsilon} \propto \varepsilon^{1/2}$, which exhibits that the frequency splitting scales proportionally to the square root of the perturbation strength. Such square root behavior originates from the underlying EP degeneracy of the eigenfrequencies. As shown in Fig. 3c, we extract frequency splitting from the experimental data, which is obviously proportional to $\varepsilon^{1/2}$ and more sensitive than conventional linear response sensors.

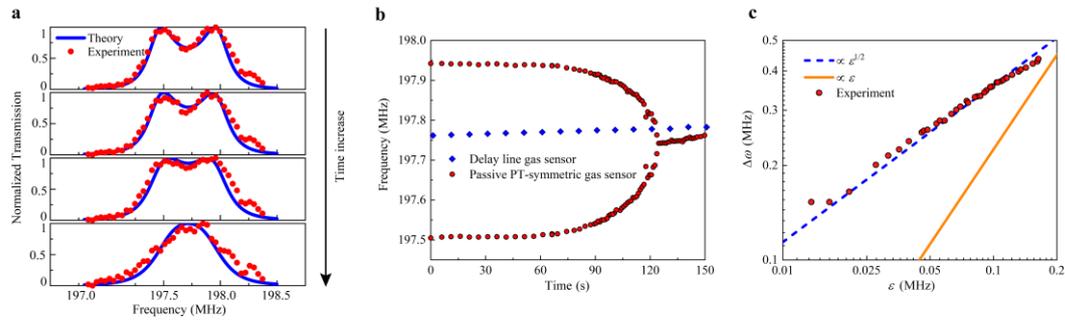

**Fig. 3 | Experimentally measured response of the sensor. a**, Transmission spectrum of our passive PT-symmetric SAW gas sensor as time increase. Red dots represent the experimental data and blue curves is the numerical fitting by employing the coupled mode theory. **b**, The red dots represent the measured frequencies of the transmission peaks of passive PT-symmetric SAW gas sensor. The blue squares represent the frequency shift of the conventional delay line gas sensor. **c**, The frequency splitting $\Delta\omega$ of passive PT-symmetric SAW gas sensor versus the perturbation $\varepsilon$ in a double-logarithmic scale. The blue dashed line and orange solid line indicates the square-root and linear scaling with perturbation $\varepsilon$, respectively.

**Sensing performance**

In this section, we demonstrate some sensing performance tests of our SAW gas sensors. Fig. 4a presents the average response time from split transmission peaks to single peak with different $H_2S$ gas concentration to evaluate the responsiveness of our sensor. When the concentration of $H_2S$ gas exceeds 2ppm, the response of the sensor is extremely fast (within 10 seconds), showcasing the excellent response performance of the sensors. As the concentration decreases, the average response time gradually increases. When the concentration has reached a very low level, such as 0.1ppm, the sensor still retains the ability to respond $H_2S$ gas but takes more time (over ten minutes) for the transmission peak transition to occur.

Fig. 4b shows the robustness of our sensor at different temperatures. In the PT exact phase regime, as the temperature gradually increases, the peaks in the transmission spectrum undergo a global blue shift, while the frequency splitting does not change significantly. Since our sensor is intrinsically self-referenced, the sensing relies on the measurement of frequency difference between two peaks, then the global frequency drift caused by temperature variation (or other external sources) will not lead to misjudgment in gas detection. But the conventional SAW delay line gas sensor which relies on the measurement of the individual frequency shift does not possess this advantage. For example, as the increase of gas concentration or the temperature, the transmission peaks will wholly blue shift[41,42] and they have similar individual frequency shift. Then it is hard to judge that the frequency shift is originated from the growth of detected gas or merely due to the increase of environment temperature.

$SnO_2$ film is a universal sensing material to exclusively respond to $H_2S$ gas, that makes our sensor having excellent selectivity to detect $H_2S$ in a mixed gas, which is distinctly demonstrated in Fig. 4c. We firstly slowly continuously inject a reducing gas $NH_3$ into the reaction chamber. After about 300s, although the concentration of $NH_3$ gas reach about 1500ppm, the transmission spectrum almost has no change. Next, we subsequently inject an oxidizing gas $NO_2$ to reach similar concentration, the transmission spectrum still keeps unchanged. Ultimately when we inject slight $H_2S$ gas (about 0.5ppm), the sensor rapidly responds and the two peaks quickly coalesce to be one in about 90s. After discharging the mixed gas, the sensor gradually recovers to initial state, that demonstrates the excellent recovery ability of our sensor.

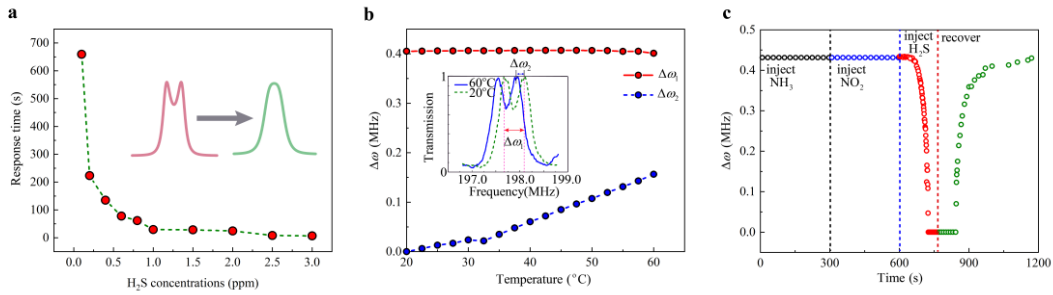

**Fig. 4 | The performance of sensor. a**, In environment with different concentrations of $H_2S$ gas, the average reaction time of sensor from two split transmission peaks to single peak. **b**, The global frequency shift of two peaks (blue solid dots, $\Delta\omega_2$) and the frequency splitting of transmission peaks (red solid dots, $\Delta\omega_1$) at different temperatures. Insert visually reveals the global frequency drift phenomenon at 20°C and 60°C. **c**, Response of sensor to the mixed gas with sequentially injecting $NH_3$ (black circle), $NO_2$ (blue circle) and $H_2S$ (red circle). And the sensor recovers to the initial state as the evacuation of gas (green circle).

**Conclusions**

Loss, which leads to energy dissipation and the decrease of the quality factor in sensor device, is usually recognized as a drawback and is minimized as much as possible. However, in this work we positively utilize the additional loss to obtain ultrasensitive sensing performance. The introduction of the additional loss in a two coupled SAW resonators structure forms a passive PT-symmetric system. Even a tiny amount of $H_2S$ gas is injected, our device exhibits a process in which two peaks merge into single peak in the transmission spectrum. When the system is in the PT exact phase regime and approaches the coalescence point of two transmission peaks, their frequency difference is proportional to the square root of perturbation strength, which is much more sensitive than the conventional linear response delay line sensor. Our sensor also presents outstanding robustness against temperature variation and exclusiveness to other gases. Our approach can be easily extended to design other types of sensors and implement other functions, such as wireless sensor, the detection of other gases or a more sensitive sensor based on higher-order EP. In addition, SAW system provides an excellent platform to realize non-Hermitian system. Besides loss, SAW system is also convenient to tune the resonant frequency and coupling strength. It is feasible to manufacture diverse non-Hermitian devices for the exploration of intriguing and rich non-Hermitian physics.

## Method

### Experimental set-ups and measurement

Our SAW chip is composed of IDTs, Bragg mirrors and thin film fabricated on a 128° Y-cut $LiNbO_3$ substrate. Both IDTs and Bragg mirrors are composed of 100nm thick and 4.9μm wide aluminum electrodes, and the pitch of both is 9.8μm. The number of electrodes of input and output IDTs is 60. The Bragg mirrors are shorted-circuited gratings and the numbers of electrodes for the Bragg mirrors near the input port, in the middle and near the output port are 40, 60 and 50, respectively. The lengths of two SAW resonators are 500μm and 402μm. A 200nm thick, 300μm width $SnO_2$ thin film is deposited onto the surface of the resonator near the input port by using magnetron sputtering technique. In the fabricating process, the base pressure to deposition is below $9 \times 10^{-4}$ Pa and during depositing it is kept at 2Pa. The sputtering power is set at 100W. The device with deposited thin film is annealed at 773.15K for 2 hours.

Our experiment setup (Fig.1c) is composed of a gas reaction chamber (with wall thickness 1cm, size $1dm \times 1dm \times 1dm$, a lid on the top of the reaction chamber that can be opened, two valves for injecting and evacuating gas, and some electrodes for connection), the prepared SAW chip (being wire-bonded to the printed circuit board) and a network analyzer. In experiment, we seal the reaction chamber and inject a predetermined amount of gas. A LabVIEW program controls the network analyzer to measure the $S_{21}$ (transmissivity equals to $|S_{21}|^2$) of the SAW device at the time cycle of one second. When the transmission spectrum no longer changes with time, we open the air valve and use an air pump to inject air and exhaust the detected gas. After the detected gas is released, the device returns to its initial state and can be used to carry out the next experiment. The sensing performance tests of conventional delay line sensor adopts the same method.

**Simulation**

We utilize COMSOL Multiphysics to simulate the experiment based on the 2D cross-section module of SAW device. We apply solid mechanics module and electrostatics module, and consider the couple of strain field and electrostatic field due to the piezoelectric effects. The whole device except the top surface is surrounded by perfectly matched layers to absorb the radiated waves. The Bragg mirrors are set as floating potential. We use mechanical damping module and dielectric loss module of piezoelectric substrate in area where the thin film has been deposited to simulate the additional loss from the thin film.

**Acknowledgements**

This work was supported by National Key R&D Program of China under Grant 2022YFE0103300 and 2020YFA0211400.

**Author contributions**

X. L. performed analytic derivation and numerical simulations, conceived and designed the experiments. F. C. fabricated the SAW devices. Y. Y. contributed method to deposit film. X. L., Y. G. and X. H. performed the experiments. X. L. wrote the manuscript. L. R. helped to revise and comment the manuscript. D. Z. and W. L. conceived the idea and supervised the project.

method and its application as sensing films of the resistance and SAW $H_2S$ sensor. *Sens. Actuators B: Chem.* **217**, 119-128 (2015).